%% file: main.tex
\documentclass[journal]{IEEEtran}
\usepackage{amsmath,amsfonts,graphicx,tabularx,color}
\usepackage{amssymb,textcomp,mathtools,amsthm,gensymb}
\usepackage{bm,upgreek,algorithm,hyperref}
\usepackage{multirow,booktabs,hhline,array,caption,subcaption}
\usepackage{cite,url,makecell,setspace,listings,verbatim}
\RequirePackage{doi}

\definecolor{codegreen}{rgb}{0,0.6,0}
\definecolor{codegray}{rgb}{0.5,0.5,0.5}
\definecolor{codepurple}{rgb}{0.58,0,0.82}
\definecolor{backcolour}{rgb}{0.95,0.95,0.92}

\renewcommand{\vec}[1]{\bm{\mathrm{#1}}}

\theoremstyle{definition}

\theoremstyle{plain}

\theoremstyle{remark}

\title{Gull: A Generative Multifunctional Audio Codec}

\author{Yi~Luo, Jianwei~Yu, Hangting~Chen, Rongzhi~Gu, Chao~Weng}

\begin{document}
\maketitle

\begin{abstract}
\input{0_abstract}
\end{abstract}

\begin{IEEEkeywords}
Neural audio codec, band-split RNN
\end{IEEEkeywords}

\section{Introduction}
\label{sec:intro}
\input{1_introduction}


\section{Gull Codec}
\label{sec:gull}
\input{3_gull}

\section{Experiment configurations}
\label{sec:config}
\input{4_config}

\section{Results and analysis}
\label{sec:result}
\input{5_result}

\section{Conclusion and Future Works}
\label{sec:conclusion}
\input{6_conclusion}

\bibliographystyle{IEEEbib}
\bibliography{refs}

\end{document}

%% file: 0_abstract.tex
We introduce Gull, a \textit{\textbf{g}}enerative m\textit{\textbf{ul}}tifunctiona\textit{\textbf{l}} audio codec. Gull is a general purpose neural audio compression and decompression model which can be applied to a wide range of tasks and applications such as real-time communication, audio super-resolution, and codec language models. The key components of Gull include (1) universal-sample-rate modeling via subband modeling schemes motivated by recent progress in audio source separation, (2) gain-shape representations motivated by traditional audio codecs, (3) improved residual vector quantization modules, (4) elastic decoder network that enables user-defined model size and complexity during inference time, (5) built-in ability for audio super-resolution without the increase of bitrate. We compare Gull with existing traditional and neural audio codecs and show that Gull is able to achieve on par or better performance across various sample rates, bitrates and model complexities in both subjective and objective evaluation metrics. A demo page is available online\footnote{\url{https://yluo42.github.io/Gull}}.

%% file: 1_introduction.tex
Audio codecs play an important role in data storage and transmission and daily communication by compressing audio waveforms into representations at a lower bitrate and decompressing them to reconstruct the original input. Traditional audio codecs were typically designed based on carefully and cleverly handcrafted audio features, advanced signal processing techniques and psychoacoustic models to jointly consider the signal and perceptual qualities \cite{atal1970adaptive, schroeder1985code, bessette2002adaptive}, which provide high quality with an adequate bitrate but introduce distortions and artifacts at lower bitrates. In recent years, neural networks have been applied to audio codecs to either enhance the output of traditional audio codecs \cite{biswas2020audio, hwang2021enhancement, lin2022speech}, improve the reconstruction quality at very low bitrates \cite{kleijn2018wavenet, klejsa2019high, garbacea2019low, kleijn2021generative, kim2021neurally}, or to perform end-to-end compression and decompression \cite{kankanahalli2018end, zeghidour2021soundstream, lim2022end, defossez2022high, kumar2023high}. Such neural audio codecs are typically \textit{generative}, i.e., trained with adversarial loss \cite{goodfellow2020generative}, aiming to address the tradeoff between signal distortion and perceptual quality by approaching the perception-distortion bound \cite{blau2019rethinking}.

The core building blocks for both traditional and neural audio codecs are encoder, quantizer and decoder. With the development of more and more advanced neural networks, handcrafted encoder features and conventional signal transformations have been gradually replaced by learnable representations via different neural network architectures, and the quantizer has been modified to support joint optimization with the rest two parts. Nevertheless, the improved quality of neural audio codecs at lower bitrates often comes with the cost of model complexity, and a single codec model is generally trained to support audio at a fixed, predefined sample rate (e.g., 16 kHz for a most common case for speech). In this case, various models need to be trained for different audio sample rates, which greatly increases the training cost. Moreover, traditional audio codecs may support different algorithmic complexities with different reconstruction qualities \cite{rfc6716}, but almost all existing neural codecs do not have such property, as the encoder and decoder are fixed once trained. In order to deploy the neural codec into platforms with different computational cost requirements, the codec typically needs to be trained at the minimal computational cost across all target platforms, which limits the modeling power of the neural network model and sacrifices the codec's performance on high-end platforms.

\begin{figure*}[!ht]
  \small
  \centering
  \includegraphics[width=2\columnwidth]{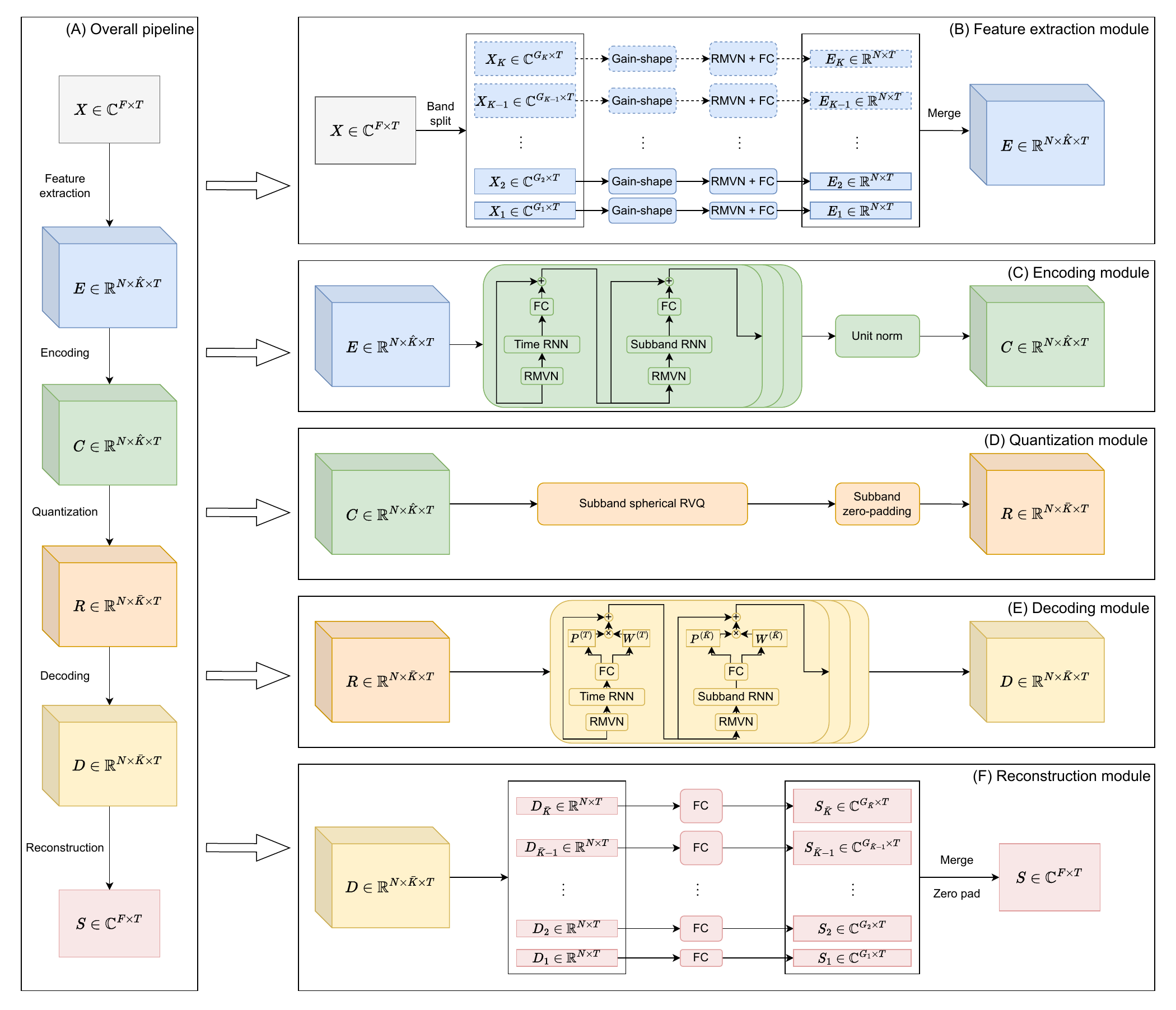}
  \caption{Flowchart of the model architecture of Gull codec.}
  \label{fig:gull}
\end{figure*}

To address the aforementioned issues, we propose \textit{\textbf{Gull}}, a \textit{\textbf{g}}enerative m\textit{\textbf{ul}}tifunctiona\textit{\textbf{l}} neural audio codec operates in the frequency domain, as the first fully-neural audio codec that supports \textit{universal sample rate, dynamic bitrate and dynamic complexity} modeling of audio signals, thanks to the recent progress in band-split audio modeling for audio source separation \cite{luo2023music, yu2023efficient}. Gull splits the input spectrogram into subbands and extracts the \textit{gain-shape representations} for each subband feature at each frame, and performs localized encoding via stacked residual feed forward networks (FFNs). Motivated by the widely-used residual vector quantization (RVQ) method \cite{zeghidour2021soundstream, lee2022autoregressive, defossez2022high, kumar2023high}, the feature quantization process is done by a spherical residual vector quantization (SRVQ) module where a unit-norm feature is quantized by a unit-norm codevector followed by series of \textit{rotation matrices}. The decoder contains band-split neural network layers with dynamic width to support universal sample rate and dynamic complexity processing, and one single model can be applied to audio with any sample rate and deployed to platforms with different computational cost requirements. Adversarial training is applied to balance the signal distortion and perceptual quality of the reconstructed signal, and it also enables Gull's built-in bandwidth extension property \cite{li2021real}. Experiments are designed to compare Gull with traditional audio codecs and recent neural audio codecs on both subjective and objective metrics, and we show that Gull is able to obtain on par or better performance than other systems at various signal sample rates, bitrates and model complexities with a theoretical system latency of 20~ms.

The rest of the paper is organized as follows. Section~\ref{sec:gull} provides a detailed description of the proposed Gull codec. Section~\ref{sec:config} provides the configurations for training and evaluation. Section~\ref{sec:result} presents the experiment results and analysis. Section~\ref{sec:conclusion} concludes the paper.

%% file: 3_gull.tex
Figure~\ref{fig:gull} shows the flowchart of the model architecture of the proposed Gull codec. Gull operates at the time-frequency domain and applies subband-level feature extraction, encoding, quantization and decoding. The feature extraction module calculates subband embeddings via \textit{gain-shape representations} for each subband, and the encoder contains stacked band-split RNN (BSRNN) blocks \cite{luo2023music} to perform joint temporal and subband modeling. The quantizer is a modified RVQ module with subband-specific codebooks. The decoder contains stacked elastic BSRNN blocks \cite{li2023subnetwork}, which supports input waveforms of various sample rates \cite{yu2023efficient}, user-defined model size and complexity, and optional super-resolution ability. The reconstruction module generates reconstructed complex-valued spectrogram from the decoder outputs.

\subsection{Signal transformation}

In order for Gull codec to support signals of various sample rates, the input signal is first resampled to the \textit{operating sample rate} of Gull. If the actual sample rate is lower than the operating sample rate, Gull only processes the valid frequency bands and ignores the others.

\subsection{Feature extraction}
\label{sec:feature}

Given the complex-valued spectrogram of the resampled input waveform $\vec{X}\in\mathbb{C}^{F\times T}$ where $F$ and $T$ denote the number of frequency bins and frames, respectively, we first split $\vec{X}$ into $K\geq1$ subband spectrograms $\vec{X}_k \in \mathbb{C}^{G_k\times T}, k=1,\ldots, K$, where $\sum_{k=1}^K G_k = F$. Motivated by pyramid vector quantization (PVQ) \cite{valin2009high}, we extract the \textit{gain-shape representation} $\vec{g}_{k,t} \in \mathbb{R}^{2G_k+1}$ for subband $k$ at frame $t$:
\begin{align}
\begin{split}
    \vec{g}_{k,t} = \text{concat}[&\text{Re}(\vec{x}_{k,t})/||\vec{x}_{k,t}||_2, \\
    &\text{Im}(\vec{x}_{k,t})/||\vec{x}_{k,t}||_2, \\
    &\text{log}(||\vec{x}_{k,t}||_2)]
\end{split}
\end{align}
where $\text{Re}(\cdot), \text{Im}(\cdot)$ represent the real and imaginary parts of a complex number, respectively, and $||\cdot||_2$ denotes the L2 norm. Unlike several existing traditional codecs where the energy is explicitly encoded by additional bandwidth or neural codecs where the energy is implicitly encoded via the encoder and the input signals are typically normalized to a certain energy range \cite{defossez2022high}, the gain-shape representation here decouples the content (shape) and energy (gain) of the subband spectrograms and forces the encoder to learn proper mappings to preserve both of them. We empirically observe that by using such gain-shape representation, input energy normalization is no longer needed and the codec is able to successfully reconstruct signals with a wide range of energy levels.

$\vec{g}_{k,t}$ is then sent to subband-specific normalization and fully-connected (FC) layers to map to $N$-dimensional embeddings $\vec{e}_{k,t} \in \mathbb{R}^{N}$, where the normalization module is a rescaled mean-variance normalization (RMVN) layer similar to frame-level layer normalization \cite{ba2016layer, luo2019conv}:
\begin{align}
    \text{RMVN}(\vec{g}_{k,t}, \vec{\alpha}_k, \vec{\beta}_k) = \frac{\vec{g}_{k,t} - \text{E}[\vec{g}_{k,t}]}{\sqrt{\text{Var}[\vec{g}_{k,t}] + \epsilon}} \odot \vec{\alpha_k} + \vec{\beta}_k
\end{align}
where $\vec{\alpha}_k, \vec{\beta}_k \in \mathbb{R}^{2G_k+1}$ are subband-specific trainable rescaling factors and $\epsilon$ is added for numerical stability. The embeddings from the subbands that covers the valid frequency bands are merged to form $\vec{E} \in \mathbb{R}^{N\times \hat{K}\times T}$, where $\hat{K}\leq K$ denotes the number of valid subbands corresponds to the input sample rate.

\subsection{Encoding}
\label{sec:encoding}

We use stacked BSRNN blocks \cite{luo2023music} in the encoder to perform joint temporal and subband modeling. A BSRNN block contains two tandem residual RNN layers sequentially scanned  across the temporal and subband dimensions, where both RNN layers are set causal. A causal RNN layer across subband dimension ensures that the modeling of lower bands is not affected by the information in higher bands, which enables the encoder to generate identical features at lower bands with different input sample rates. The output of the last BSRNN block, $\vec{C}\in\mathbb{R}^{N\times \hat{K}\times T}$, is normalized along the embedding dimension $N$ to have unit L2 norm.

\subsection{Quantization}
\label{sec:quantization}

To better quantize the unit-norm embeddings $\vec{c}_{k,t}\in\mathbb{R}^{1\times N}$, we modify the standard RVQ module to spherical RVQ (SRVQ) module. For a SRVQ module with $H$ hierarchies, we set the first hierarchy to a standard VQ layer with unit-norm codebooks, and the other hierarchies to contain learnable rotation matrices defined by Householder transform. To be more specific, for the first hierarchy, the code selection process is defined as:
\begin{align}
\begin{split}
    \vec{e}^1_{k,t} &= \vec{q}^1_{k,j}, \,j =
    \text{argmin}_{i} ||\vec{q}^1_{k,i} - \vec{c}_{k,t}||_2 \\
\end{split}
\end{align}
where $\vec{q}^1_{k,j}\in\mathbb{R}^{1\times N}$ denotes the $j$-th code in the codebook at subband $k$. Starting from the $h$-th hierarchy where $h\geq2$, the code selection process is defined as:
\begin{align}
\begin{split}
    \vec{e}^h_{k,t} &= \vec{e}^{h-1}_{k,t}\vec{O}^h_{k,j}, \,j =
    \text{argmin}_{i} ||\vec{e}^{h-1}_{k,t}\vec{O}^h_{k,j} - \vec{c}_{k,t}||_2 \\
\end{split}
\end{align}
where $\vec{O}^h_{k,j}\in\mathbb{R}^{N\times N}$ is a rotation matrix\footnote{Here we treat reflection matrices as rotation matrices for similar effect.} calculated by Householder transformation with unit-norm vector $\vec{o}^h_{k,j}\in\mathbb{R}^{1\times N}$:
\begin{align}
    \vec{O}^h_{k,j} = \vec{I} - 2\vec{o}^{h\top}_{k,j}\vec{o}^h_{k,j}
\end{align}
$\vec{o}^h_{k,j}$ is a learnable parameter jointly optimized with the rest of the system. To ensure that the quantization error is non-increasing, we set $\vec{o}^h_{k,0} =\vec{0}$ to ensure that there is always an identity rotation matrix (i.e., identity mapping). 

During training, we apply the exponential moving average (EMA) update \cite{razavi2019generating} to the codebooks corresponding to valid input subbands at the first hierarchy, and we randomly replace codes that have not been used in 100 consecutive iterations by a random embedding in $\vec{c}_{k}$ at the current iteration. We also randomly select a hierarchy $h\leq H$ similar to existing works \cite{defossez2022high, kumar2023high} to allow the decoder to handle embeddings quantized at different bitrates. For the other hierarchies, we apply the standard VQ commitment loss to update the rotation embeddings $\vec{o}^h_{k,j}$. The overall quantization commitment loss is calculated at all hierarchies for the $\hat{K}$ valid subbands:
\begin{align}
    l_{\text{commit}} =& \frac{1}{\hat{K}T}\sum_{k=1}^{\hat{K}} \sum_{t=1}^T ||\vec{c}_{k,t} - sg[\vec{e}^1_{k,t}]||_2 + \\
    & \frac{1}{\hat{K}T(H-1)}\sum_{k=1}^{\hat{K}} \sum_{t=1}^T \sum_{h=2}^H ||\vec{c}_{k,t} - sg[\vec{e}^h_{k,t}]||_2 + \\
    & \frac{1}{\hat{K}T(H-1)}\sum_{k=1}^{\hat{K}} \sum_{t=1}^T 
    \sum_{h=2}^H||sg[\vec{c}_{k,t}] - \vec{e}^h_{k,t}||_2
\end{align}
where \textit{sg} denotes the stop-gradient operation. 

We denote the final quantized embedding tensor as $\vec{R}^h \in \mathbb{R}^{N\times\hat{K}\times T}$. To support optional audio super-resolution property, we can set an output sample rate with its corresponding number of valid subbands $\bar{K}\geq\hat{K}$, and perform zero-padding between the $\hat{K}$-th and $\bar{K}$-th subbands to augment $\vec{R}^h$ to shape $\mathbb{R}^{N\times\bar{K}\times T}$.

\begin{figure*}[!ht]
  \small
  \centering
  \includegraphics[width=2\columnwidth]{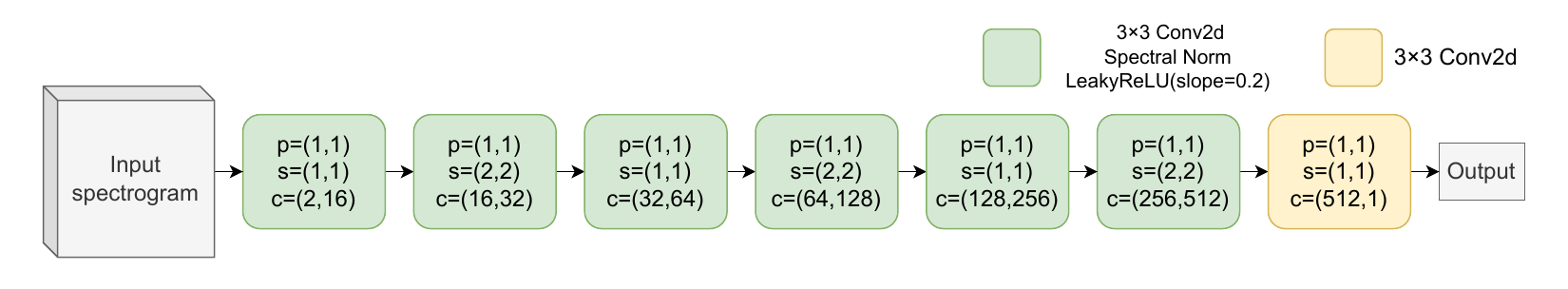}
  \caption{Flowchart of the STFT discriminator.}
  \label{fig:GAN}
\end{figure*}

\subsection{Decoding}
\label{sec:decoding}

The quantized embedding $\vec{R}^h$ is passed to stacked elastic BSRNN blocks \cite{li2023subnetwork} to generate decoded embeddings $\vec{D}\in\mathbb{R}^{N\times \bar{K}\times T}$. An elastic BSRNN block contains two interleaved residual elastic RNN layers, where one is applied across the temporal dimension $T$ and one is applied across the valid subband dimension $\bar{K}$. Each residual elastic RNN layer contains an RMVN module for frame-level input normalization, an RNN layer to perform temporal or subband modeling, and an FC layer that generates outputs and weighting scores of different network ``width''. For an RNN output frame $\vec{z}_{k,t} \in \mathbb{R}^{M}$, the FC layer generates $\vec{p}_{k,t}\in\mathbb{R}^{N\times W}$ and $\vec{d}_{k,t}\in\mathbb{R}^{V\times W}$, where $W$ denotes the ``width'' of the layer. $\vec{d}_{k,t}$ is further sent to a transform-average-concatenate (TAC) \cite{luo2020end} module to generate weighting scores $\vec{w}\in\mathbb{R}^{1\times H}$ given a selected layer width $w\leq W$:
\begin{align}
\begin{split}
    \vec{u}_{k,t,:w} &= \text{Tanh}(\text{FC}(\vec{d}_{k,t,:w})) \\
    \vec{q}_{k,t} &= \text{Tanh}(\text{FC}(\frac{1}{w}\sum_{i=1}^w \vec{u}_{k,t,i})) \\
    \vec{b}_{k,t,i} &= \text{FC}(\text{concat}[\vec{u}_{k,t,i}, \vec{q}_{k,t}]),\,i=1,\ldots, w \\
    \vec{w}_{k,t,:w} &= \text{Softmax}_{w}(\vec{b}_{k,t,:w})
\end{split}
\end{align}
where $\vec{u}_{k,t,:w}\in\mathbb{R}^{N\times w}$ denotes the first $w$ embeddings of $\vec{u}_{k,t}$. The final output is obtained by $\sum_{i=1}^w \vec{p}_{k,t,i}\cdot \vec{w}_{k,t,i}$. It is easy to find that $\vec{q}_{k,t}$ and $\vec{w}_{k,t,:w}$ are width-dependent, which means that for different selected layer widths $w$ the weighting scores of the first $w$ features $\vec{w}_{k,t,:w}$ are different. This defines the elastic property of the layer. Similarly, for a selected network depth $d\leq D$ where $D$ denotes the total number of elastic BSRNN layers in the decoder, we use the first $d$ layers to generate the decoder output and omit the remaining layers. This is equivalent to the early-exit mechanism \cite{teerapittayanon2016branchynet} and allows the model to support dynamic-depth inference. We denote the final output of the decoder as $\vec{d}_{k,t}^{w,d}\in\mathbb{R}^{N}$ with width $w$ and depth $d$.

\subsection{Reconstruction}
\label{sec:reconstruct}

$\vec{d}_{k,t}^{w,d}$ is passed to subband-specific FC layers to generate the real and imaginary parts of the reconstructed subband spectrogram ${\vec{s}_{k,t}^{w,d}\in\mathbb{C}^{G_k}}$. We use gated linear unit (GLU) \cite{dauphin2017language} as the nonlinearity for the FC layer. The $\bar{K}$ reconstructed subband spectrograms are concatenated along the frequency dimension, and we perform zero-padding at the high-frequency bands when $\bar{K}<K$. The final reconstructed complex-valued spectrogram $\vec{S}^{w,d}\in\mathbb{C}^{F\times T}$ is passed to inverse short-time Fourier transform (ISTFT) to convert back to waveform at the operating sample rate, and another optional resampling process can be applied to transform the waveform back to the actual sample rate of the input signal.

\subsection{Adversarial training}
\label{sec:GAN}

We utilize multi-resolution STFT discriminators similar to existing works \cite{defossez2022high} for adversarial training. Figure~\ref{fig:GAN} shows the model architecture for the STFT discriminator, which contains stacked 2D convolutional blocks with leaky ReLU activation \cite{maas2013rectifier} and spectral normalization \cite{miyato2018spectral}. $p$ and $s$ denote the padding and stride sizes, respectively, and $c$ represents the input and output channels of the block. The input to the discriminators is the valid frequency bands of the input or reconstructed spectrogram, where the real and imaginary components are stacked to form the 3D tensor with channel dimension of 2. We normalize the signals to have unit L2 norm before sending them into the discriminators to ensure that the discriminators are insensitive to input energy.

\subsection{Discussion}

Traditional audio codecs such as MP3 \cite{rfc3119}, AAC \cite{bosi1997iso}, Opus \cite{rfc6716} and EVS \cite{dietz2015overview} made use of advanced signal transformation and coding techniques such as modified discrete cosine transform (MDCT), linear predictive coding (LPC), vector quantization, range coding, and subband modeling. Neural networks were mainly used as an auxiliary module, e.g., for voice activity detection (VAD). Built upon conventional audio coding techniques, neural networks were designed to lower the bitrate while enhancing the performance \cite{kleijn2018wavenet, klejsa2019high, biswas2020audio, kim2021neurally}. Fully-neural audio codecs elevate the position of neural networks by using them instead of conventional signal processing algorithms to transform, encode and decode the audio waveforms, and use either scalar-wise or bin-wise quantization \cite{kankanahalli2018end, zhen2021scalable, yang2021source, petermann2021harp, lim2022end, lim2023end, kim2023progressive} or various forms of vector quantization techniques \cite{garbacea2019low, kleijn2021generative, zeghidour2021soundstream, chen2021tenc, lee2022progressive, defossez2022high, jenrungrot2023lmcodec, kumar2023high} to quantize and dequantize the intermediate representations. However, none of the existing fully-neural audio codecs support dynamic audio sample rates or dynamic decoding complexities. Gull, together with the universal-sample-rate band-split modeling scheme, is the first neural audio codec that holds such properties without the need of a pre-configured traditional audio codec.

The feature extraction module splits the input spectrogram into subbands. For tasks such as codec language models where codes are treated as discrete representations for training a language model \cite{borsos2023audiolm, wang2023neural, wang2023speechx, copet2023simple, agostinelli2023musiclm}, one can either use subband Gull codes with corresponding modifications to the language models or reconfigure Gull to perform full-band modeling to serve as a direct alternative to existing full-band codecs.

The decoder described in section~\ref{sec:decoding} makes use of autoregressive RNN layers for temporal and subband modeling. However, any other autoregressive or nonautoregressive network architectures can be easily adopted to save computational complexity, model size or support faster decoding. For example, a noncausal temporal convolutional network (TCN) \cite{luo2019conv} can replace the temporal RNN layer to allow parallel decoding.

\begin{table*}[!ht]
  \caption{Model statistics with different configurations. ``SR'' stands for sample rate and ``comp.'' stands for complexity.}
  \label{tab:config}
  \small
  \centering
  \scalebox{0.99}{
  \begin{tabular}{c|c|c|c|c|c|c|c}
    \toprule
    \small
    Model & Input & Target & \multirow{2}{*}{Bitrate (kbps)} & \multirow{2}{*}{$w$} & \multirow{2}{*}{$d$} & Encoder & Decoder comp. \\
    type & SR (kHz) & SR (kHz) & & & & comp. (MACs/s) & (MACs/s, w.r.t target SRs) \\
    \midrule
    \multirow{15}{*}{Speech} & \multirow{3}{*}{8} & \multirow{3}{*}{8/16/24/32/48} & \multirow{3}{*}{1.2×\{2-6\}} & 1 & 1 & \multirow{3}{*}{176.5M} & 69.3M / 138.6M / 207.9M / 277.2M / 348.6M \\
     & & & & 1 & 4 & & 271.0M / 542.1M / 813.1M / 1.1G / 1.4G \\
      & & & & 10 & 4 & & 498.6M / 997.2M / 1.5G / 2.0G / 2.5G \\
      \cline{2-8}
     & \multirow{3}{*}{16} & \multirow{3}{*}{16/24/32/48} & \multirow{3}{*}{2.4×\{2-6\}} & 1 & 1 & \multirow{3}{*}{352.9M} & 138.6M / 207.9M / 277.2M / 348.6M \\
     & & & & 1 & 4 & & 542.1M / 813.1M / 1.1G / 1.4G \\
      & & & & 10 & 4 & & 997.2M / 1.5G / 2.0G / 2.5G \\
     \cline{2-8}
    & \multirow{3}{*}{24} & \multirow{3}{*}{24/32/48} & \multirow{3}{*}{3.6×\{2-6\}} & 1 & 1 & \multirow{3}{*}{529.4M} & 207.9M / 277.2M / 348.6M \\
     & & & & 1 & 4 & & 813.1M / 1.1G / 1.4G \\
      & & & & 10 & 4 & & 1.5G / 2.0G / 2.5G \\
     \cline{2-8}
    & \multirow{3}{*}{32} & \multirow{3}{*}{32/48} & \multirow{3}{*}{4.8×\{2-6\}} & 1 & 1 & \multirow{3}{*}{705.8M} & 277.2M / 348.6M \\
     & & & & 1 & 4 & & 1.1G / 1.4G \\
      & & & & 10 & 4 & & 2.0G / 2.5G \\
     \cline{2-8}
    & \multirow{3}{*}{48} & \multirow{3}{*}{48} & \multirow{3}{*}{6.0×\{2-6\}} & 1 & 1 & \multirow{3}{*}{883.3M} & 348.6M \\
     & & & & 1 & 4 & & 1.4G \\
      & & & & 10 & 4 & & 2.5G \\
    \hline
    \multirow{12}{*}{Music} & \multirow{3}{*}{16} & \multirow{3}{*}{16/24/32/44.1} & \multirow{3}{*}{8.4×\{2-6\}} & 1 & 1 & \multirow{3}{*}{1.2G} & 474.8M / 544.1M / 613.4M / 683.8M \\
     & & & & 1 & 4 & & 1.9G / 2.2G / 2.4G / 2.7G \\
      & & & & 10 & 4 & & 3.5G / 4.0G / 4.5G / 5.0G \\
     \cline{2-8}
    & \multirow{3}{*}{24} & \multirow{3}{*}{24/32/44.1} & \multirow{3}{*}{9.6×\{2-6\}} & 1 & 1 & \multirow{3}{*}{1.4G} & 544.1M / 613.4M / 683.8M \\
     & & & & 1 & 4 & & 2.2G / 2.4G / 2.7G \\
      & & & & 10 & 4 & & 4.0G / 4.5G / 5.0G \\
     \cline{2-8}
    & \multirow{3}{*}{32} & \multirow{3}{*}{32/44.1} & \multirow{3}{*}{10.8×\{2-6\}} & 1 & 1 & \multirow{3}{*}{1.6G} & 613.4M / 683.8M \\
     & & & & 1 & 4 & & 2.4G / 2.7G \\
      & & & & 10 & 4 & & 4.5G / 5.0G \\
    \cline{2-8}
    & \multirow{3}{*}{44.1} & \multirow{3}{*}{44.1} & \multirow{3}{*}{12.0×\{2-6\}} & 1 & 1 & \multirow{3}{*}{1.8G} & 683.8M \\
     & & & & 1 & 4 & & 2.7G \\
      & & & & 10 & 4 & & 5.0G \\
    \bottomrule
  \end{tabular}
  }
\end{table*}

%% file: 4_config.tex
\subsection{Dataset}

We train Gull codec on speech and music data to evaluate its performance on different types of signals. For speech model, we use the English-only clean 48 kHz speech from the 5th deep noise suppression (DNS) challenge  \cite{dubey2022icassp}, which contains 391.7 hours of data. For music model, we use an internal dataset of 570.7 hours of 44.1 kHz lossless music. All audio are segmented into 3-second chunks for model training, and no energy adjustment was made in any data. For evaluation, we use a 1-hour subset randomly selected from the VCTK dataset \cite{veaux2017cstr} and the Expresso dataset \cite{nguyen2023expresso} for the speech model, and a 1-hour internal dataset of lossless music and sound effect for the music model. None of the evaluation data are included in the training set.

\subsection{Hyperparameters}

We use slightly different hyperparameter settings for the speech model and the music model. For speech model, the operating sample rate is set to 48 kHz, and we split the input spectrogram into 10 bands, where the first 8 bands have 2 kHz bandwidth and the last 2 bands have 4 kHz bandwidth. For music model, the operating sample rate is set to 44.1 kHz, and we split the input spectrogram into 20 bands, which contain ten 400 Hz subbands, four 1 kHz subbands, four 2 kHz subbands, one 4 kHz subband, and the rest as a whole. The window and hop sizes for short-time Fourier transform (STFT) for calculating the spectrograms are 20~ms and 10~ms, respectively, which result in a theoretical system latency of 20~ms. The embedding dimension $N$ is set to 64 in both models, and the number of encoder and decoder layers is set to 4. We use LSTM layers in the decoder, and the hidden dimension is set to 128. The width of each LSTM layer (i.e., $W$) is set to 10, the dimension of FC layer in the RNN module (i.e., $N+V$ in section~\ref{sec:decoding}) is set to 80 ($N=64, V=16$), and the hidden dimension in the TAC module is set to 16. For dynamic bandwidth support, we use $H=5$ hierarchies of codebooks in the SRVQ modules with 12 bits in the first hierarchy and 6 bits in the rest. Table~\ref{tab:config} shows the supported bitrates and actual model complexity (MACs per second) for different input and target sample rates and 3 selected decoder widths and depths. Note that no entropy coding techniques are applied to the SRVQ codes, hence the bitrates reported here can be further optimized with other external modules or algorithms. For adversarial training, we use 5 discriminators with the STFT window sizes of 256, 512, 1024, 2048 and 4096, respectively, and the hop sizes are set to half of the window sizes.

All models are trained with 4 NVIDIA V100 GPUs for 200 epochs with 5000 iterations per epoch. We use AdamW \cite{loshchilov2018decoupled} optimizer with an initial learning rate of 0.001 for the codec and 0.0001 for the discriminators, and the learning rate is decayed by 0.98 for every 2 epochs. For each training sample in each iteration, we randomly sample an input sample rate and a target sample rate, where for speech the input sample rate is sampled from 8/16/24/32/48 kHz and for music the input sample rate is sampled from 16/24/32/44.1 kHz. The target sample rate is sampled to be greater than or equal to the input sample rate. As mentioned in section~\ref{sec:gull}, the input and target sample rates define the valid and target subbands, respectively.

We select Encodec \cite{defossez2022high}, the state-of-the-art (SOTA) neural audio codec which supports both speech and music, as a benchmark neural codec system to compare with. For a fair comparison, we use the official model implementation\footnote{https://github.com/facebookresearch/encodec} with causal configuration but train it with the same data and pipeline. For speech model, we only train it for 16 kHz sample rate with the downsampling factors for the encoding layers set to (5,4,4,2), and for music model, we only train it for 44.1 kHz sample rate with the downsampling factors of (7,7,3,3). Both of them contain 100 latent steps per second of audio, which match the configuration of Gull. We also use 5 residual blocks for RVQ, the speech model supports 2.4×\{2-6\} kbps bitrates, and the music model supports 12.0×\{2-6\} kbps bitrates. The model complexity for the speech and music models are 3.9G and 8.1G MACs, respectively.

\subsection{Training objectives}

The discriminators are updated by standard least-square GAN (LSGAN) loss \cite{mao2017least}:
\begin{align}
\begin{split}
    l_{\text{LS},i}^{w,d} &= \text{E}[(D_i(\vec{X}) - 1)^2] + \text{E}[D_i(\vec{S}^{w,d})^2] \\
    l_{\text{Dis}} &= \frac{1}{5} \sum_{i=1}^5 (l_{\text{LS},i}^{w,d} + l_{\text{LS},i}^{W,D})
\end{split}
\end{align}
where $w, d$ denote the randomly sampled decoder width and depth, $W, D$ denote the full decoder width and depth, and $D_i$ denotes the $i$-th discriminator.

The generator (i.e., codec) loss consists of a reconstruction loss, a feature matching loss, a discriminator loss, and a VQ-VAE commitment loss. The reconstruction loss operates at multiple STFT resolutions, and it includes the mean absolute error (MAE) between the magnitude spectrograms and the mel-frequency spectrograms of the reconstructed and target signals:
\begin{align}
\begin{split}
    l_{\text{Freq}}^{w,d} &= \text{E}[\frac{||\vec{S}^{w,d}| - |\vec{X}||}{\text{mean}(|\vec{X}|)}] \\
    l_{\text{Mel}}^{w,d} &= \text{E}[\frac{|\text{Mel}(\vec{S}^{w,d}) - \text{Mel}(\vec{X})|}{\text{mean}(\text{Mel}(\vec{X}))}] \\
    l_{\text{Rec}} &= l_{\text{Freq}}^{w,d} + l_{\text{Freq}}^{W,D} + l_{\text{Mel}}^{w,d} + l_{\text{Mel}}^{W,D}
\end{split}
\end{align}
Following recent works \cite{kumar2023high}, we set the STFT window sizes to 32, 64, 128, 256, 512, 1024, 2048, and set the corresponding number of mel-frequency bands to 5, 10, 20, 40, 80, 160, 320.

The feature matching loss is defined as the layer-wise normalized MAE between all discriminator hidden representations of the reconstructed and target signals, normalized across the 6 layers in the 5 discriminators:
\begin{align}
\begin{split}
    l_{\text{Feat}, i}^{w,d} &= \frac{1}{6} \sum_{j=1}^6 \text{E}[\frac{|\vec{FS}_{i,j}^{w,d} - sg[\vec{FX}_{i,j}]|}{\text{mean}(|sg[\vec{FX}_{i,j}]|)}] \\
    l_{\text{FM}} &= \frac{1}{5} \sum_{i=1}^5 (l_{\text{Feat},i}^{w,d} + l_{\text{Feat},i}^{W,D})
\end{split}
\end{align}
where $\vec{FS}_{i,j}, \vec{FX}_{i,j}$ denote the $i$-th discriminator's hidden representations at $j$-th layer for reconstructed and target signals, respectively.

The discriminator loss also follows the standard update scheme for LSGAN, averaged across 5 discriminators:
\begin{align}
\begin{split}
    l_{\text{DG},i}^{w,d} &= \text{E}[D_i(\vec{S}^{w,d})^2] \\
    l_{\text{Gen}} &= \frac{1}{5} \sum_{i=1}^5 (l_{\text{DG},i}^{w,d} + l_{\text{DG},i}^{W,D})
\end{split}
\end{align}

The total generator loss is then defined as:
\begin{align}
    l_{\text{total}} = l_{\text{Rec}} + l_{\text{FM}} + l_{\text{Gen}} + \alpha\cdot l_{\text{commit}}
\end{align}
where we empirically set $\alpha=0.2$.

\subsection{Evaluation metrics}

For objective metrics, we use signal-to-noise ratio (SNR), perceptual evaluation of speech quality (PESQ) \cite{rix2001perceptual}, and virtual speech quality objective listener (VISQOL)\footnote{https://github.com/google/visqol} \cite{chinen2020visqol}. For subjective metric, we design a MUSHRA test \cite{series2014method} similar to prior works \cite{defossez2022high} but without the low-pass anchor. 30  subjects are recruited to rate the perceptual quality of the audio samples in a range between 0 to 100, and we randomly select 15 samples from both evaluation sets to serve as the audio samples. All subjects are required to wear headphones and perform the test in a quiet environment. Subjects who rate the hidden reference sample below 90 for more than 15 percent of all test samples are set disqualified.

%% file: 5_result.tex
\begin{figure*}
     \centering
     \begin{subfigure}[b]{2\columnwidth}
         \centering
         \includegraphics[width=\columnwidth]{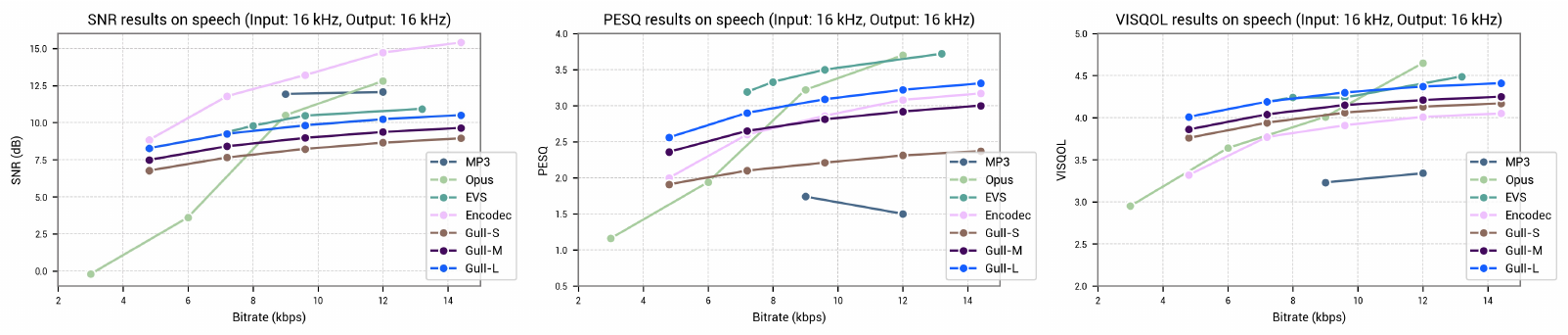}
         \caption{PESQ, SNR and VISQOL results for various audio codec configurations on 16 kHz speech.}
         \label{fig:result-speech-16}
     \end{subfigure}
     \hfill
     \begin{subfigure}[b]{2\columnwidth}
         \centering
         \includegraphics[width=\columnwidth]{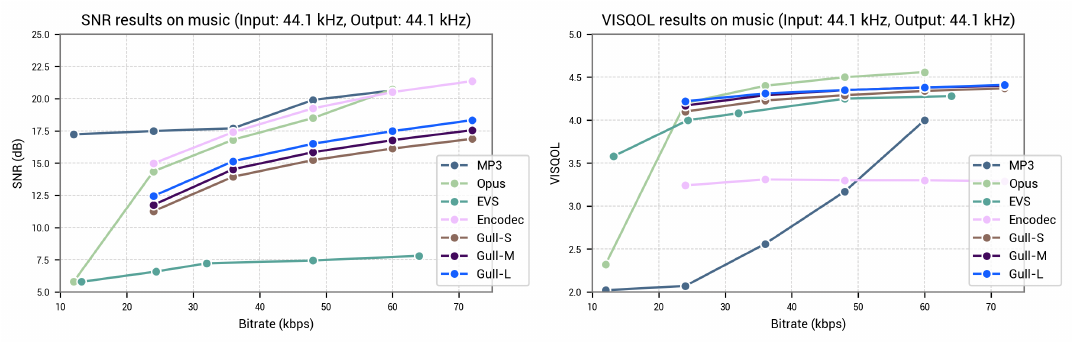}
         \caption{SNR and VISQOL results for various audio codec configurations on 44.1 kHz music.}
         \label{fig:result-music-441}
     \end{subfigure}
     \hfill
     \begin{subfigure}[b]{2\columnwidth}
         \centering
         \includegraphics[width=\columnwidth]{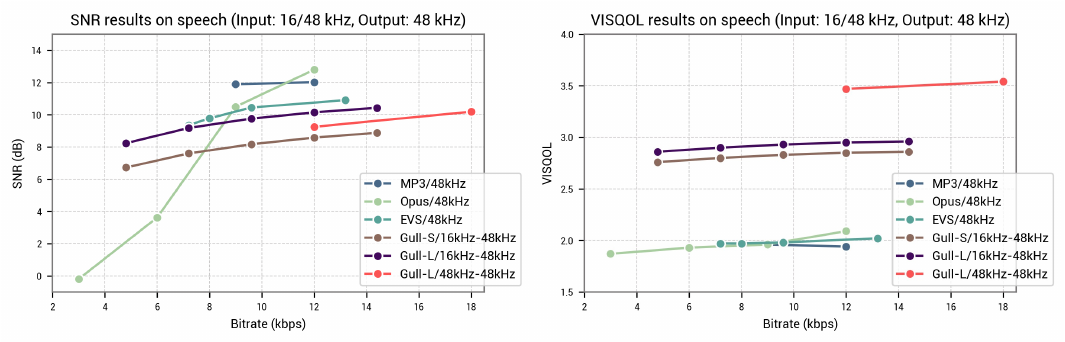}
         \caption{SNR and VISQOL results for various Gull configurations for joint codec and bandwidth extension on speech.}
         \label{fig:result-speech-sr}
     \end{subfigure}
     \begin{subfigure}[b]{2\columnwidth}
         \centering
         \includegraphics[width=\columnwidth]{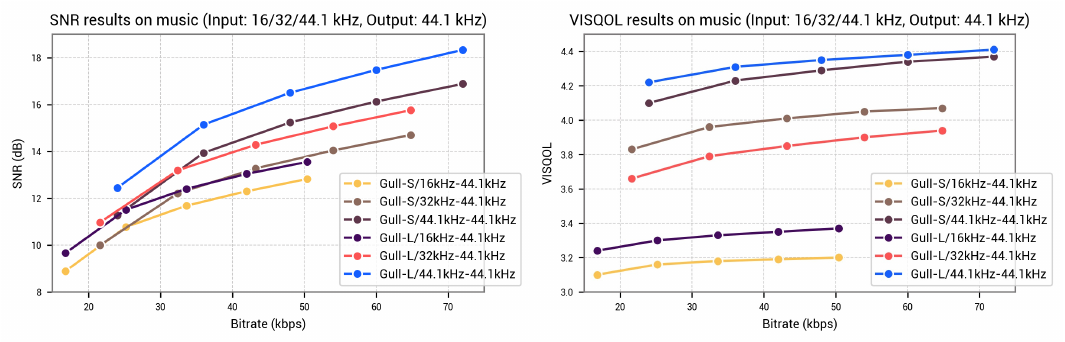}
         \caption{SNR and VISQOL results for various Gull configurations for joint codec and bandwidth extension on music.}
         \label{fig:result-music-sr}
     \end{subfigure}
     
    \caption{Subjective evaluation results for Gull and other benchmark conventional and neural audio codecs.}
    \label{fig:subjective-results}
\end{figure*}

\begin{table*}[!ht]
  \caption{MUSHRA scores for different codecs under various bandwiths and complexities. We report mean scores and 95\% confidence intervals. Speech signals are sampled at 16 kHz, and music signals are sampled at 44.1 kHz.}
  \label{tab:mos}
  \small
  \centering
  \begin{tabular}{c|ccc|ccc}
    \toprule
    \small
    \multirow{2}{*}{Codec} & \multicolumn{3}{c|}{Speech} & \multicolumn{3}{c}{Music} \\
    & Bitrate (kbps) & comp. (MACs/s) & MUSHRA & Bitrate (kbps) & comp. (MACs/s) & MUSHRA \\
    \midrule
    Reference & -- & -- & 92.4\scriptsize{±0.1} & -- & -- & 92.8\scriptsize{±0.2} \\
    \cline{1-7}
    MP3 & 6.0 & -- & 64.4\scriptsize{±1.0} & 24.0 & -- & 71.0\scriptsize{±1.6} \\
    \cline{1-7}
    Opus & 6.0 & -- & 59.2\scriptsize{±0.9} & 24.0 & -- & 92.3\scriptsize{±0.3} \\
    \cline{1-7}
    EVS & 7.2 & -- & 85.7\scriptsize{±2.4} & 24.4 & -- & 91.8\scriptsize{±0.4} \\
    \cline{1-7}
    Encodec & 4.8 & 3.9G & 61.6\scriptsize{±1.0} & 24.0 & 8.1G & 83.7\scriptsize{±1.0} \\
    \cline{1-7}
    \multirow{3}{*}{Gull-S} & 2.4 & 315.1M & 52.8\scriptsize{±0.8} & 16.8 & 1.9G & 85.8\scriptsize{±0.9} \\
     & 3.6 & 315.1M & 53.7\scriptsize{±0.8} & 21.6 & 2.3G & 89.7\scriptsize{±0.6} \\
     & 4.8 & 491.5M & 62.9\scriptsize{±1.0} & 24.0 & 2.4G & 90.2\scriptsize{±0.5} \\
    \cline{1-7}
    \multirow{3}{*}{Gull-L} & 2.4 & 1.2G & 75.6\scriptsize{±0.9} & 16.8 & 6.2G & 91.7\scriptsize{±0.4} \\
    & 3.6 & 1.2G & 80.1\scriptsize{±0.8} & 21.6 & 6.6G & 90.7\scriptsize{±0.5} \\
    & 4.8 & 1.4G & 84.1\scriptsize{±0.5} & 24.0 & 6.7G & 92.5\scriptsize{±0.3} \\
    \bottomrule
  \end{tabular}
\end{table*}

Figure~\ref{fig:subjective-results} shows the objective scores for various codecs on speech and music test sets. Due to the space limit and the vast amount of possible input/output sample rates and decoder model sizes of Gull, here we only select three Gull complexities, Gull-S, Gull-M and Gull-L, corresponding to decoder configurations of $\{w=1,d=1\}$, $\{w=1,d=4\}$ and $\{w=10,d=4\}$, respectively, and a few selected standard input/output sample rate configurations. Figure~\ref{fig:result-speech-16} and~\ref{fig:result-music-441} present the evaluation of standard codec ability across different bitrates and different Gull model sizes, and Figure~\ref{fig:result-speech-sr} and~\ref{fig:result-music-sr} present the evaluation of joint codec and bandwidth extension ability of different Gull input/output sample rates and model sizes. 

For 16 kHz speech data, we can see from Figure~\ref{fig:result-speech-16} that the PESQ and VISQOL scores of Gull with different model sizes are consistently better than Opus at lower bitrates ($\leq$ 10 kbps) and MP3 at all bitrates, showing that the model capacity as well as the reconstruction ability of the first few SRVQ layers are able to effectively model the most important contents (PESQ) and maintain a reasonable sound quality (VISQOL). With a similar level of model complexity, Gull-L outperforms Encodec across all bitrates on PESQ and VISQOL, showing that the proposed model architecture and SRVQ module might have higher capacity on better quantizing and dequantizing the input signals. For middle to high bitrates, EVS performs better than all other codecs on PESQ, possibly due to its highly optimized codec ability to wideband speech. Gull-L can achieve almost on par VISQOL scores as EVS below 12 kbps. Among all codecs, Encodec always has the highest SNR, possibly due to the time-domain network better reconstructs the phase information. 

For 44.1 kHz music data, we can see from Figure~\ref{fig:result-music-441} that Gull configurations have worse SNR scores but much higher VISQOL scores than MP3 and Encodec. Opus performs consistently better than all other codecs on VISQOL with $\geq$ 20 kbps, while EVS performs slightly worse that Gull across all bitrates. Interestingly, our reproduced Encodec model performs bad on VISQOL across all bitrates, and one possible explanation might be that the bit assignment as well as the training paradigm in its RVQ module can become inefficient at higher bitrates (more codebooks), which limits its performance when the bitrate is high. As an opposite, subband-level quantization scheme in Gull allows us to flexibly find the balance between the number of subbands, number of bits in each subband and the number of RVQ levels, which bypasses the problem of assigning too many bits (i.e., codebook sizes) in one RVQ layer.

For the joint codec and bandwidth extension task for speech data, we select the input sample rate to 16 kHz and the output sample rate to 48 kHz and calculate the objective scores. For music data, we set the input sample rate to 16 or 32 kHz and the output sample rate to 44.1 kHz. The VISQOL scores for both test sets are calculated at 48 kHz sample rate. Figure~\ref{fig:result-speech-sr} and ~\ref{fig:result-music-sr} show that Gull can use fewer bits for transmission and leverage the power of the decoder to generate higher sample rate outputs from lower sample rate inputs, and such process can lead to improved objective scores compared to the codec-only processing scheme.

Table~\ref{tab:mos} presents the subjective ratings of different codecs measured by MUSHRA scores. The Gull models with 2.4 kbps and 3.2 kbps for speech signals are the ones with 8 kHz inputs sample rate but bandwidth extended to 16 kHz output sample rate with 1 or 2 RVQ layers, and the models with 16.8 kbps and 21.6 kbps for music signals are the ones with 16 kHz and 32 kHz input sample rates but bandwidth extended to 44.1 kHz output sample rate with 1 RVQ layer. We can see that Gull is able to achieve on-par or better performance than conventional codecs at same or lower bitrates, and an increased model complexity consistently improves the MUSHRA score. Compared to Encodec, Gull is able to achieve higher user ratings with the small model and significantly better performance with the large model, whose model size and complexity are still lower than the selected Encodec configuration.

%% file: 6_conclusion.tex
In this paper, we proposed Gull, a neural audio codec for universal-sample-rate audio codec with optional bandwidth extension ability. Gull utilized several recent progresses on subband modeling neural networks for audio front-end processing, residual vector quantization techniques for audio and image tokenization, and adversarial training schemes for perceptual-enhanced data reconstruction and generation. Objective and subjective evaluation results showed that compared to various other conventional codecs and a strong neural audio codec, Gull was able to achieve on par or better performance in both measurements with smaller model sizes or bitrates.

Future works for Gull include various aspects. First, matching the computational constraints for real-time communication (RTC) scenarios still requires extra effort on optimizing the model size and complexity of the systems. Second, how to effectively apply Gull to audio generation frameworks with either language models \cite{wang2023speechx} or flow-based or diffusion-based models \cite{shen2023naturalspeech, tan2024naturalspeech} also needs further investigation. Third, as the band-split model architecture used in Gull has already proven effective in various speech and audio front-end processing tasks \cite{luo2023music, yu2023tspeech, yu2023efficient}, performing end-to-end audio codec, enhancement/separation and bandwidth extension is a natural direction to explore.